# Modeling of the drag and thermal conductivity coefficients in tropical cyclones taking into account foam cover.

Ephim Golbraikh,[1] and Yuri M. Shtemler,[2]

[1]*Department of Physics, Ben-Gurion University of the Negev, P.O. Box 653, Beer-Sheva 84105, Israel*

[2]*Department of Mechanical Engineering, Ben-Gurion University of the Negev, P.O. Box 653, Beer-Sheva 84105, Israel*

**Abstract**

A formal averaging procedure over the air-sea interface is developed for both momentum and enthalpy surface-transfer coefficients, $C_D$ and $C_K$, in hurricane conditions. This leads to splitting of both the transfer coefficients across the total area of the sea surface, $C_{D,K}$, into the sums of their partial values over the foam-free, $C_{D,Kw}$, and foam-covered, $C_{D,Kf}$, fractions weighted with foam- and water-coverage coefficients $α_f$ and 1-$α_f$. The transfer coefficients, $C_{D,K}$ and $C_{D,Kw}$, are estimated by measurements in open-sea and in laboratory conditions, respectively, while the transfer coefficients across the foam-covered fraction, $C_{D,Kf}$, which cannot be measured directly, are estimated from the splitting relations. Applying the Monin-Obukhov similarity theory at the neutral stability atmospheric conditions to the transfer coefficients, separately to the foam-free, the foam-covered, and the total sea surfaces, yields the roughness lengths, $Z_{D,Kw}, Z_{D,Kf}, Z_{D,K}$. The study is aimed at explaining an anomalous behavior of the momentum and enthalpy transfer coefficients, $C_D$ and $C_K$, with wind speed $U_{10}$ under hurricane conditions, by the effect of the foam slipping layer sandwiched between the atmosphere and the ocean.

## 1. Introduction

The present paper is devoted to the study of the surface momentum and enthalpy transfer coefficients, $C_D$ and $C_K$, varied with the hurricane wind speed $U_{10}$ at a reference height 10 [$m$]. The momentum transfer from the atmosphere to the ocean in hurricane conditions has been numerously studied (e.g., Powell et al. 2003; Donelan et al. 2004; Black et al. 2007; Jarosz et al. 2007; Shtemler et al. 2010; Holthuijsen et al. 2012; Soloviev et al. 2014; Golbraikh and Shtemler 2016; Golbraikh and Shtemler 2018, thereafter GS2016; GS2018, and references therein). Concerning similar studies of the heat transfer through the air-sea interface in hurricane conditions, note that they have been investigated much less intensively than the momentum transfer. The laboratory measurements demonstrate that foam coverage significantly affects heat transfer across the interface (Nekrasov *et al*. 1988; Chickadel 2015). However, from these results it is difficult to draw conclusions about heat transfer under hurricane conditions. In addition, both drag and enthalpy transfer coefficients, $C_D$ and $C_K$, are measured with significant errors. Below, the heat transfer under hurricane conditions is investigated basing on only a few available data for the open sea (e.g. Zhang et al. 2008; Bell et al. 2012; Richter et al. 2016) and laboratory (Jeong et al. 2012; Sergeev et al. 2017; Komori et al. 2018) measurements.

The measurements made in laboratory conditions demonstrate that $C_D$ monotonically increases with $U_{10}$ up to its saturation at $U_{10} \approx 35\ [ms^{-1}]$, while $C_K$ shows a nearly-constant value up to $U_{10} \approx 35\ [ms^{-1}]$, and then starts a monotonic quasi-linear growth with $U_{10}$ (see dashed lines depicted by $C_D$ and $C_K$ vs $U_{10}$ in Fig. 1 and Fig. 4, respectively). According to the available laboratory measurement data, this more or less regular behavior of the transfer coefficients in $U_{10}$ is changed into a rather unexpected one in the open-sea hurricane conditions. Thus, $C_D$ increases up to a maximum at $U_{10} \approx 35\ [ms^{-1}]$, then reaches a minimum at $U_{10} \approx 55\ [ms^{-1}]$, after which it starts to increase again with $U_{10}$. Simultaneously, $C_K$ monotonically decreases up to $U_{10} \approx 45\ [ms^{-1}]$, then has a nearly-constant value up to $U_{10} \approx 55\ [ms^{-1}]$, and after that begins to grow again (see solid lines depicting $C_D$ and $C_K$ vs $U_{10}$ in Fig. 1 and Fig. 4, respectively). The references corresponding to the dashed and solid lines are also presented in the figure captions, and omitted here for brevity. The anomalous behavior of both momentum and heat transfer coefficients with $U_{10}$ does not have a commonly accepted explanation yet.

There are different scenarios of these anomalies mainly concerning the surface drag coefficient under hurricane conditions. A short recent review of these scenarios of the aerodynamic drag saturation/reduction with growing hurricane winds can be found in Troitskaya et al. 2019 (see also references therein, and in addition Zhao et al. 2015; Bye et al. 2014; Andreas and Mahrt 2016; Obermann et al. 2016; Sergeev et al. 2017; Zhao and Li 2018; Donelan 2018). Roughly speaking, these scenarios can be separated into three groups which explore either (i) different wave effects (such as flow separation from breaking wave crests, Kelvin-Helmholtz instability of the air-sea interface, etc.), (ii) effect of spray, or (iii) the slipping effect of the foam layer sandwiched between the atmosphere and the ocean.

The present study concerns the effect of the foam slipping layer covering the atmosphere-ocean interface on the momentum and heat transfer across it in hurricane conditions (Shtemler et al. 2010; Holthuijsen et al., 2012; GS2016; GS2018; Takagaki et al. 2016; MacMahan 2017; Troitskaya et al. 2019). Below, our recent investigations of the drag transfer in hurricane conditions (GS2016; GS2018) are generalized and developed for accounting the heat transfer effects. The study is aimed at explaining the anomalous behavior of the momentum and enthalpy transfer coefficients, $C_D$ and $C_K$, with wind speed $U_{10}$ due to the effect of the foam slipping layer sandwiched between the atmosphere and the ocean. With this goal, a formal averaging procedure over the interface is developed for both $C_D$ and $C_K$.

This procedure leads to splitting formulas for both transfer coefficients across the total area of the sea surfaces, $C_D$ and $C_K$, into the sums of their partial values over the foam-free, $C_{Dw}$ and $C_{Kw}$, and foam-covered, $C_{Df}$ and $C_{Kf}$, fractions weighted with foam- and water-coverage coefficients, $α_f$ and $1-α_f$. Transfer coefficients across the total area, $C_D$ and $C_K$, are estimated using open-sea measurements. Whereas transfer coefficients across the foam-free fraction, $C_{Dw}$ and $C_{Kw}$, are assumed to be measured in laboratory conditions within the interval $U_{10}$ from low to hurricane values, when the foam quantity is negligibly small, since the foam is mainly removed from the working section of the laboratory equipment (see discussion in GS2018). In this connection, note some results reported by Troitskaya et al. 2019 for the foam influence on the behavior of $C_D$ with $U_{10}$ in the range of 12–38 [$ms^{-1}$]. Their work is first directed toward laboratory investigation of the foam impact on momentum exchange in the atmospheric boundary layer. However, they note that the artificial foam produced by the foam generator may differ from the natural sea foam which arises from wave breaking in the real open-sea conditions. Finally, note that transfer coefficients across the foam-covered fraction of the interface, $C_{Df}$ and $C_{Kf}$, cannot be directly measured within the range of $U_{10}$ from low to hurricane values. Indeed, the air-water interface totally covered with foam can be only observed at extremely high $U_{10}$, when the foam coverage is close to unity. Whilst at low and intermediate values $U_{10}$, such an interface is a rather virtual object, and the splitting relations are used for estimating $C_{Df}$ and $C_{Kf}$ within the whole interval of $U_{10}$ values.

The paper is organized as follows. A physical model of the momentum and heat transfer across the atmosphere-ocean interface is presented in the next section. Summary and discussion are given in Section 3. A formal averaging procedure over the air-sea interface developed for both momentum and enthalpy transfer in hurricane conditions is presented in Appendix A. The results of the recent ad-hoc modeling of the drag transfer across the interface by GS2016, and GS2018, modified and generalized in the present paper, is shortly revised in Appendix B.

## 2. Physical model
### 2.1. Momentum transfer across the atmosphere-ocean interface
According to relation (A3) (Appendix A)

$$C_D = (1 - \alpha_f)C_{Dw} + \alpha_f C_{Df}, \tag{1}$$

where $C_D(U_{10})$ and $C_{Dw}(U_{10})$ are assumed to be measured in the open-sea and laboratory conditions, respectively, and $C_{Dw}(U_{10})$ describes the wave and spray effects, but doesn't contain the foam effects. Since $C_{Df}(U_{10})$ cannot be measured directly, the splitting relation (1) is used for their definitions.

Applying the Monin-Obukhov similarity theory at the neutral stability atmospheric conditions to the transfer coefficients, separately to the foam-free, foam-covered and total sea surfaces, yields the roughness lengths, $Z_{Dw}, Z_{Df}, Z_D$ (Eqs. (A6a), Appendix A)

$$C_{Dw} = \frac{\kappa}{ln[Z_{10}/Z_{Dw}]}, \quad C_{Df} = \frac{\kappa}{ln[Z_{10}/Z_{Df}]}, \quad C_D = \frac{\kappa}{ln[Z_{10}/Z_D]}. \tag{2}$$

In the present modeling, the measurement data for $C_D$ and $C_{Dw}$ taken as fiducial ones, are adopted from Jarosz et al. 2007 and from the modified Large-Pond model (Eq. (B3), Appendix B) for open-sea and laboratory conditions, respectively. While the modified Large-Pond model for $C_{Dw}$ is extended up to $U_{10} > 50$ [ms$^{-1}$], the reliable measurement data for $C_D$ are unavailable at such values of $U_{10}$, and the above procedure is modified as in (GS2016; GS2018). Assuming that the foam bubble sizes are characterized by the foam roughness length, $Z_{Df}$, put them constant, $Z_{Df} = Z_{Df}(50\,[{\rm ms}^{-1}])$ at $U_{10} > 50$ [ms$^{-1}$]. Then $C_{Df}$ is determined by the second relation (2), while the splitting Eq. (1) determines $C_D$ vs $U_{10}$ at $U_{10} > 50$ [ms$^{-1}$].

The resulting dependences $C_D$, $C_{Dw}$, $C_{Df}$ and $Z_D$, $Z_{Dw}$, $Z_{Df}$ vs $U_{10}$ are shown in Fig. 1 and Fig. 2, respectively. Figure 1 demonstrates that at high values of $U_{10} > 50$ [ms$^{-1}$] $C_{Dw}$ is much larger than $C_D \approx C_{Df}$. The latter reflects the fact that at extremely high wind speeds the ocean surface is almost completely covered with foam. As it is evident both GS2018 and the present model have exactly the same dependences $Z_D(U_{10})$ and $Z_{Dw}(U_{10})$, and as shown in Fig. 2, they have rather similar values of the foam roughness $Z_{Df}(U_{10})$ within the experimental error of measuring and an uncertainty under hurricane conditions. As mentioned in GS2018, $Z_{Df}$ ~$(10^{-4} \div 5 \cdot 10^{-3})$ [$m$] in Fig. 2 well correlates by the order of magnitude with the foam bubble radius, measured (e.g. Deane and Stokes 2002; Rojas and Lowelen 2007; Blenkinsopp and Chaplin 2010) or numerically simulated (Deike et al. 2016) inside breaking waves.

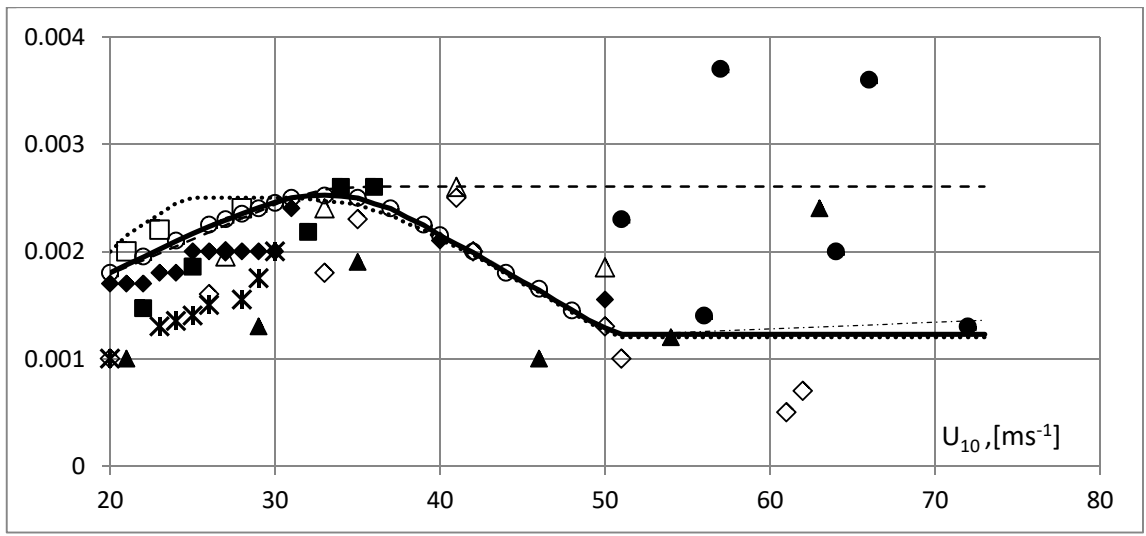

**Figure 1.** $C_D$, $C_{Dw}$, $C_{Df}$ vs $U_{10}$ [$ms^{-1}$] are solid, dashed, dotted lines.
$C_D$ vs $U_{10}$ [$ms^{-1}$] is a dashed-dotted line (with regard to a foam blowing model (7)).
Measurement points obtained in the open-sea conditions are adopted from:
Wu 1982–squares; Powell et al. 2003–triangles; Jarosz et al. 2007–circles; Holthuijsen et al. 2012–diamonds; Zhang et al. 2008–black diamonds; Bell et al. 2012 –black circles; Jeong et al. 2012–black squares; Richter et al. 2016–black triangles.

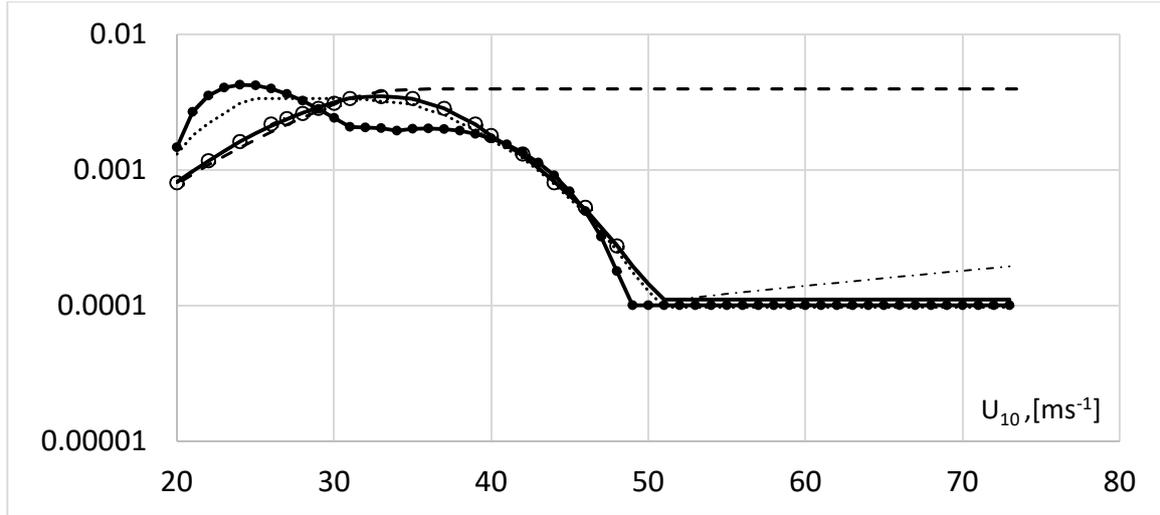

**Figure 2.** $Z_D, Z_{Dw}, Z_{Df}$ [m] vs $U_{10}$ $[ms^{-1}]$ are the solid, dashed, dotted lines. The black-circles line is $Z_{Df}$ [m] vs $U_{10}$ $[ms^{-1}]$ adopted from GS2018. Rest notations as in Fig. 1

## 2.2. Heat transfer across the atmosphere-ocean interface

The enthalpy transfer coefficient $C_K$ is treated as (see (A3), Appendix A) :

$$C_K = (1 - \alpha_f)C_{Kw} + \alpha_f C_{Kf}. \qquad (3)$$

Two enthalpy flux coefficients $C_K$ and $C_{Kw}$ are assumed to be measured in open-sea and laboratory conditions, respectively, while $C_{Kf}$ will be found from (3). Note that laboratory measurements of the enthalpy transfer coefficients in (3), as thouse of the drag transfer coefficients in (1), are carried out in foam-free conditions in the presence of both the wave and spray effects.

Figure 3 demonstrates the modeling results for $C_K$, $C_{Kw}$ and $C_{Kf}$ vs $U_{10}$ based on the splitting relation (3) and on the measurement data obtained in the open-sea (Bell et al. 2012; Richter et al. 2016), and laboratory (Komori et al. 2018) conditions. These measurement data are taken as the fiducial ones since they are only available in the range of high wind speeds $35 \ [ms^{-1}] < U_{10} < 70 \ [ms^{-1}]$. Some other measurement data for $C_K$, obtained at low wind speeds in the open-sea (Zhang et al. 2008) and in the laboratory (Jeong et al. 2012; Sergeev et al. 2017) conditions also presented in Fig. 3, are in a good agreement with the fiducial data. Besides, the results of the present modeling are in a qualitative agreement with the fact that at low wind speeds, $U_{10}$, sea foam enhances the heat flux from water by a factor of 3-5 (Chickadel 2015). Note that the sharp growth in $C_K = C_{Kw}$ in the laboratory measurements (Komori et al. 2018) at $35\ [ms^{-1}] < U_{10} < 50\ [ms^{-1}]$ makes them significantly larger than $C_K$ in the open-sea (Bell et al. 2012; Richter at al. 2016). It can be conjectured that this is related to the foam influence in the open-sea conditions as compared with laboratory ones.

The measured data for heat transfer coefficients $C_K$ and $C_{Kw}$ are nearly constant at $U_{10}$, for $U_{10} < 35[ms^{-1}]$, but they sharply vary for higher values of $U_{10}$ up to extremely high $U_{10} \sim 50\ [ms^{-1}]$, when they reach other near-constant values at $U_{10} > 50\ [ms^{-1}]$ (see Fig. 3). Note that the measurement data for $C_K$ are available within the range of $U_{10}$ from low to hurricane values, albeit with significant errors at extremely high winds. In the absence of reliable experimental data for $C_K(U_{10})$, the following plausible hypothesis for $C_{Kf}(U_{10})$ is adopted. Within the range of measurement errors, the simplest constant correlation for $C_{Kf}$ gives a sufficiently good approximation for measurement data independent of $U_{10}$ in the range of $35\ [ms^{-1}] < U_{10} < 70\ [ms^{-1}]$. The constant value $C_{Kf} \approx 0.00085$ (see Fig. 4) is obtained from Eq. (3) by using the measurement data for $C_K$ in open-sea conditions (Bell et al. 2012 and Richter et al. 2016) and for $C_{Kw}$ in laboratory conditions (Komori et al. 2018).

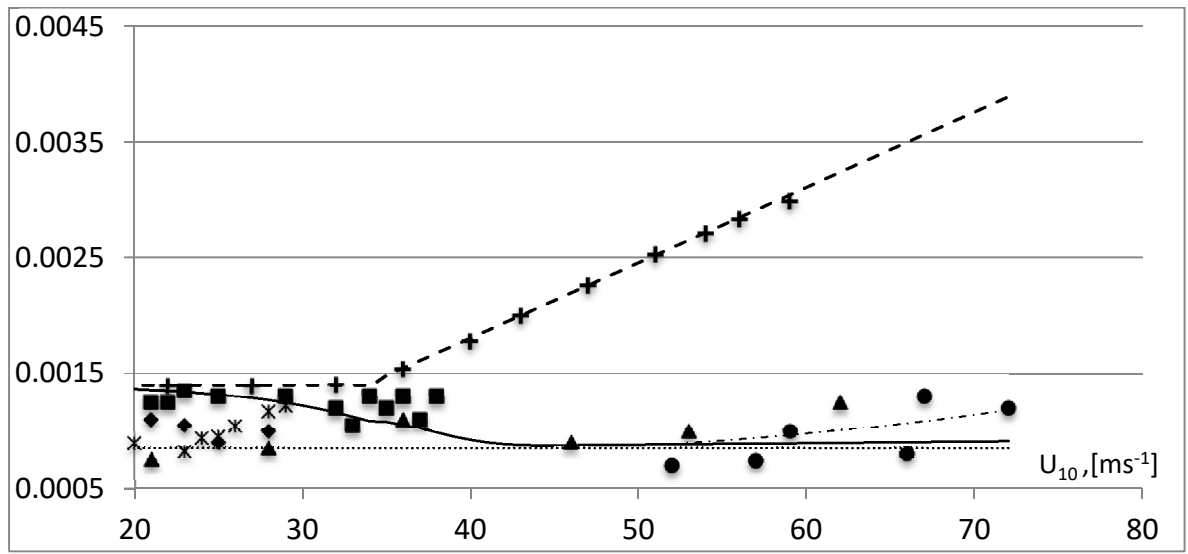

**Figure 3.** $C_K$, $C_{Kw}$, $C_{Kf}$ vs $U_{10}\ [ms^{-1}]$ are solid, dashed, dotted lines

Present modeling under the constant correlation $C_{Kf} = 0.00085$.

Notations as in Fig. 1, besides measurement points adopted from:

Sergeev et al. 2017–stars; Komori et al. 2018 - crosses (laboratory conditions).

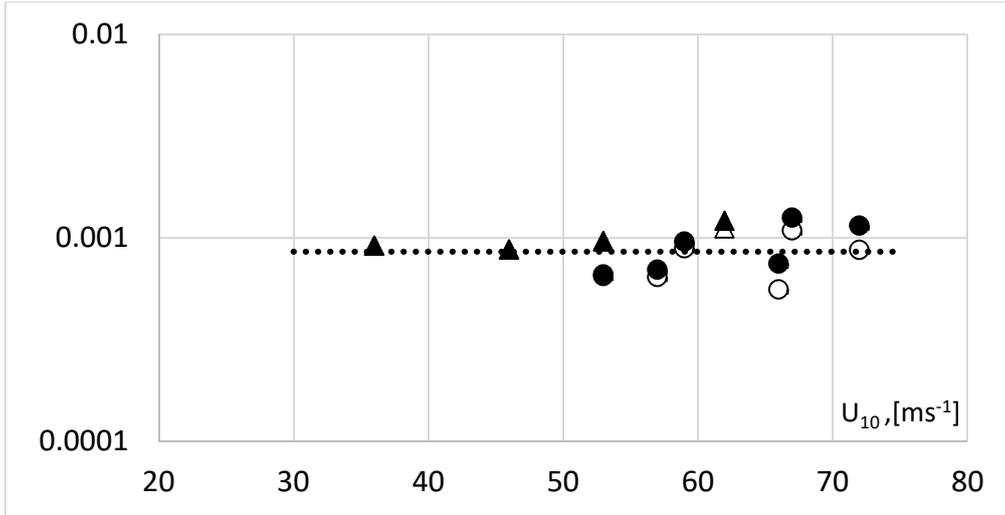

**Figure 4.** $C_{Kf}$ vs $U_{10}$ $[ms^{-1}]$ is a dotted line.
Constant correlation $C_{Kf} = 0.00085$ and $C_{Kf}$ values estimated by Eq. (3) and the fiducial measurement data for $C_K$ and $C_{Kw}$ (Bell et al. 2012; Richter at al. 2016 and Komori et al. 2018, respectively). Unfilled triangles and circles correspond to $C_{Kf}$ vs $U_{10}$ with regard to foam blowing model (7). Rest notations as in Fig. 3.

According to the analytical model of hurricanes, the square of the maximum wind speed is proportional to $C_K/C_D$, and its value lies in the range of 0.75 -1.5 (Emanuel 1995). Figure 5 demonstrates that in the current modeling $C_K/C_D$ acquires a near-constant marginal value ≈ 0.75 at $U_{10} > 50[ms^{-1}]$ in accordance with the Emanuel criterion of the real hurricane maintaining. The results of the present modeling of $C_K$ and $C_D$ show that the foam presence significantly decreases the maximum hurricane intensity as compared to foam-free laboratory conditions.

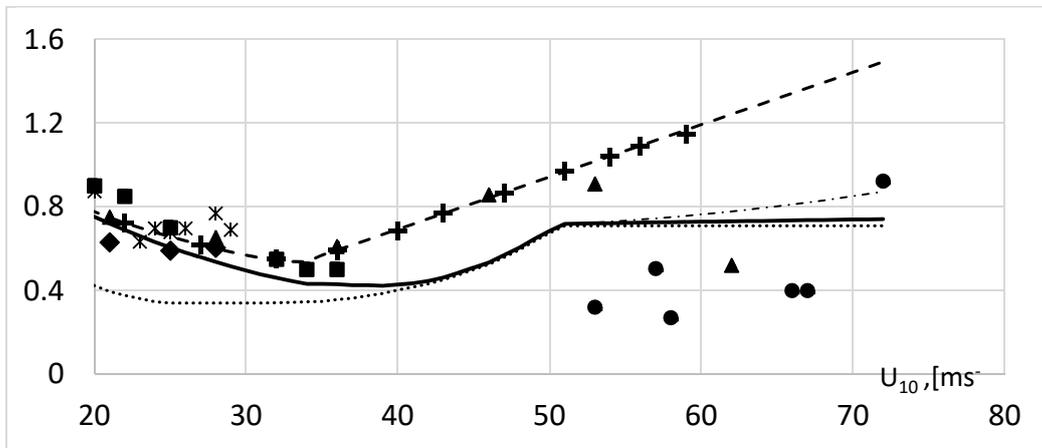

**Figure 5.** $C_K/C_D$, $C_{Kw}/C_{Dw}$, $C_{Kf}/C_{Df}$ vs $U_{10}$ $[ms^{-1}]$. Notations as in Fig. 3.

According to Monin-Obukhov similarity theory, heat transfer under neutral stability atmospheric conditions yields the following relationships between $C_K$ and $C_D$ valid for both open-sea and laboratory measurements (see Appendix A)

$$C_K = \sqrt{C_D}\frac{\kappa}{\ln[Z_{10}/Z_K]}, \tag{4}$$

$$C_{Kw} = \sqrt{C_{Dw}}\frac{\kappa}{\ln[Z_{10}/Z_{Kw}]}, \tag{5}$$

$$C_{Kf} = \sqrt{C_{Df}}\frac{\kappa}{\ln[Z_{10}/Z_{Kf}]}, \tag{6}$$

where $Z_K$, $Z_{Kw}$ and $Z_{Kf}$ are the corresponding enthalpy roughness lengths. According to Eqs. (4)-(6), it is natural to characterize the heat transfer exchange by the ratios $C_K/\sqrt{C_D}$, $C_{Kw}/\sqrt{C_{Dw}}$, $C_{Kf}/\sqrt{C_{Df}}$ which determine only the enthalpy roughness lengths (see Fig. 6).

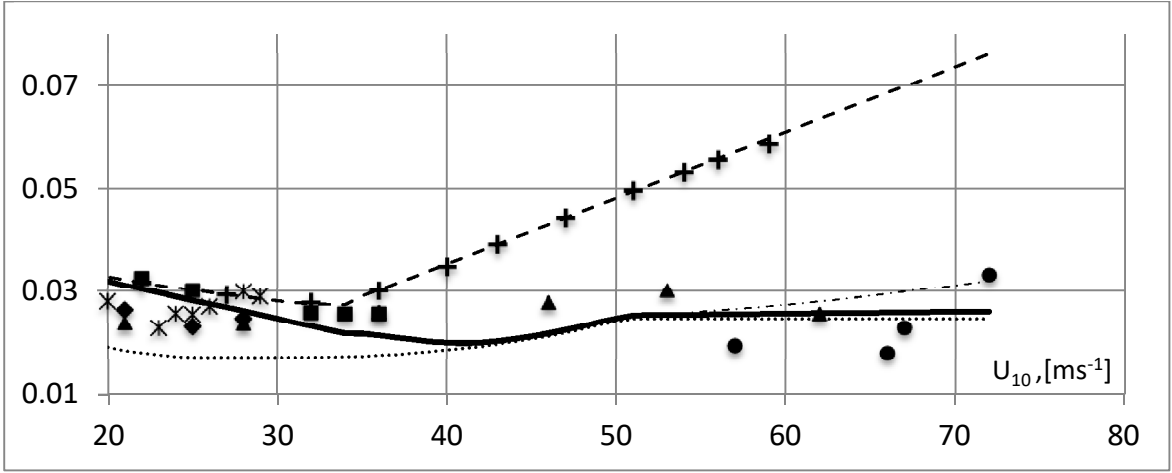

**Figure 6.** $C_K/\sqrt{C_D}$, $C_{Kw}/\sqrt{C_{Dw}}$, $C_{Kf}/\sqrt{C_{Df}}$ vs $U_{10}$ $[ms^{-1}]$. Notations as in Fig. 3.

Now $Z_K(U_{10})$, $Z_{Kw}(U_{10})$, $Z_{Kf}(U_{10})$ can be found from the relations (4)-(6), respectively. Figure 7 demonstrates the dependences $Z_K$, $Z_{Kw}$ and $Z_{Kf}$ vs $U_{10}$. Note that $Z_K(U_{10})$ varies from $Z_K \approx Z_{Kw}$ at $U_{10} \approx 20\ [ms^{-1}]$ (where $\alpha_f$ is of the order of several percents) to $Z_K \approx Z_{Kf}$ at $U_{10} \approx 43\ [ms^{-1}]$ (where $\alpha_f$ is close to unity).

Figure 7 shows that $Z_{Kw}(U_{10})$ obtained in the present modeling is in a good agreement with the results calculated on the basis of laboratory measurements (e.g., Jeong et al. 2012, Andreas and Mahrt 2016 and references therein). Note that $Z_{Kf}$ in Fig. 7 well correlates by the order of magnitude with the film thickness of the foam bubbles measured in laboratory conditions (~$10^{-9}$ - $10^{-7}$ [$m$], e.g. Exerowa et al. 2003, Yaminsky et al. 2010 and references therein).

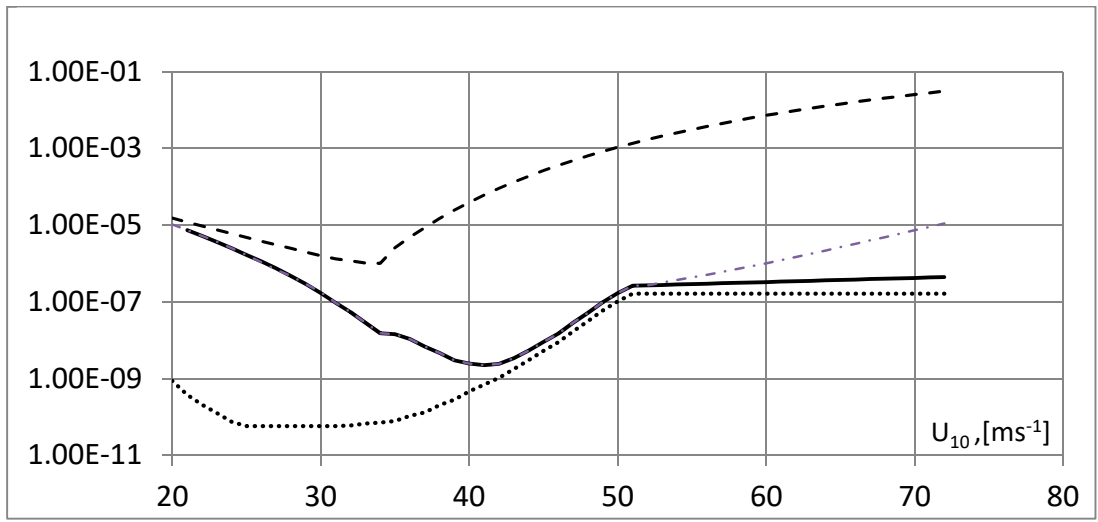

**Figure 7.** Dependence of $Z_K, Z_{Kw}, Z_{Kf}$ [$m$] vs $U_{10}$ [$ms^{-1}$]. Notations as in Fig. 3.

Figure 8 demonstrates that the enthalpy roughness length, $Z_{Kf}$, decreases when the aerodynamic roughness length, $Z_{Df}$, increases ($20\ [ms^{-1}] < U_{10} < 30\ [ms^{-1}]$) and vice versa ($35\ [ms^{-1}] < U_{10} < 50\ [ms^{-1}]$) up to $Z_{Df}$ and $Z_{Kf}$ achieve their minimal and maximal values, respectively, at large $U_{10} > 50\ [ms^{-1}]$.

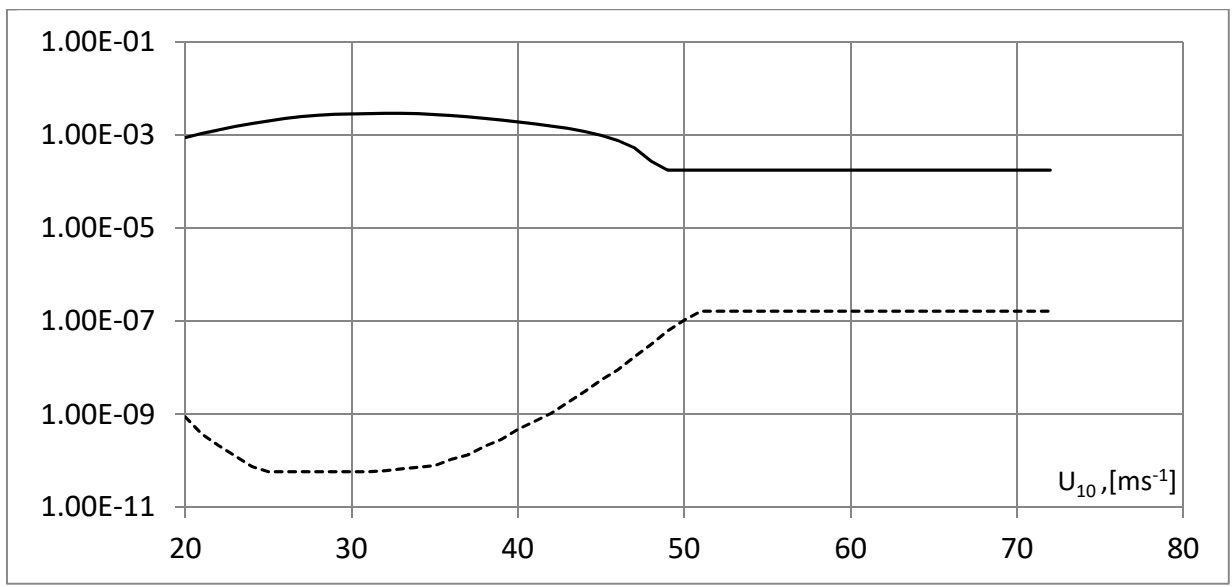

**Figure 8.** $Z_{Df}$ and $Z_{Kf}$ [$m$] vs $U_{10}$ [$ms^{-1}$] - solid and dashed lines, respectively.

## 3. Summary and Discussion

The formal averaging procedure over the interface is developed for the momentum and enthalpy transfer coefficients, $C_D$ and $C_K$. This leads to splitting of both the transfer coefficients across the total area of the alternating foam-free and foam-covered fractions of the sea surfaces, $C_D$ and $C_K$, into the sums of their partial values over the foam-free ($C_{Dw}$ and $C_{Kw}$), and foam-covered ($C_{Df}$ and $C_{Kf}$), portions weighted with foam- and water-coverage coefficients, $\alpha_f$ and 1-$\alpha_f$. According to this presentation of the transfer coefficients, all effects of kinds (i) and (ii) mentioned in the Introduction are related to $C_{Dw}$ and $C_{Kw}$, while the foam effect is related to $C_{Dw}$ and $C_{Kw}$. The foam influence on the transfer coefficients is demonstrated. It is found that at low wind speeds both $Z_{Df}$ and $Z_{Kf}$ well correlate by their order with the observed radiuses of the foam bubbles and of their water film thickness, $Z_{Df} \sim 10^{-3}$ [$m$] and $Z_{Kf} \sim 10^{-9}$ [$m$], respectively. While at large $U_{10} > 50$ [$ms^{-1}$], it may be conjectured according to the results of the present modeling that the radius of the foam bubbles and thickness of their water film are of the orders $Z_{Df} \sim 10^{-4}$ [$m$] and $Z_{Kf} \sim 10^{-7}$ [$m$], respectively.

In addition, the results of the present modeling of $C_K$ and $C_D$ show that the foam presence significantly decreases the maximum storm intensity (predicted by the Emanuel analytic model of hurricanes) in real hurricane conditions compared to the foam-free laboratory experiments. Furthermore, in foam-free laboratory experiments at extremely high wind speeds of $U_{10} > 50$ [$ms^{-1}$], the values of $C_K$ are significantly larger than those in the averaged open-sea data. It can be conjectured that this is related to the foam quantity in open-sea conditions as compared with laboratory ones.

Although the available measurement data for both aerodynamic and enthalpy exchange coefficients at high wind speed are very scarce, Figs. 1 and 4 can indicate to a significant growth of these values at $U_{10} > 50$ [$ms^{-1}$]. Such rather unexpected growth of $C_D$ and $C_K$ at high $U_{10}$ can be explained by decreasing of the foam coverage coefficient $\alpha_f$ at high values of $U_{10} > 50$ [$ms^{-1}$] because of the effect of the foam blowing from the sea surface. This effect can be modeled by the following modification of phenomenological formula (A5) for the foam coverage $\alpha_f(U_{10})$ (Appendix A) at high values of $U_{10} > 50$ [ms$^{-1}$], where there is no available data for $\alpha_f(U_{10})$:

$$\alpha_f = \begin{cases} \gamma \tanh\left[\alpha \exp\left(\beta \frac{U_{10}}{U_{10}^{(S)}}\right)\right] & \text{for } U_{10} \leq 52 \text{ [ms}^{-1}], \\ \gamma \tanh\left[\alpha \exp\left(\beta \frac{U_{10}}{U_{10}^{(S)}}\right)\right] - \delta U_{10} + \varepsilon & \text{for } U_{10} > 52 \text{ [ms}^{-1}], \end{cases} \quad (7)$$

where $\delta = 0.004347$ and $\varepsilon = 0.22$. The values of $\delta$ and $\varepsilon$ are chosen to agree the modified modeling of $C_K$ with experimental data in Figure 3. Variations arising in $C_D(U_{10})$ and $C_K(U_{10})$ calculated using (7) are shown by the dashed-dotted lines in Figs. 1 and 4-7 presented above.

Thus, the present modeling illustrates the role of foam in the slip layer formation, which determines the behavior of the momentum and heat transfer coefficients at high $U_{10}$. It also provides an explanation of possible growth of $C_D$ and $C_K$ at high $U_{10}$ by the effect of the foam blowing from the sea surface. Understanding of these processes can be useful for their implementation in numerical modeling and interpretation of the results of satellite sensing of the ocean surface and measuring its brightness temperature.

**Appendix A**

The turbulent fluxes of momentum, $\tau$, sensible heat, $Q_H$, and latent heat, $Q_E$, in the surface layer between the atmosphere and ocean are defined by bulk formulas (e.g. Garratt 1992; Andreas et al. 2012; Zou et al. 2017; Komori et al. 2018 and references therein):

$$\tau = \rho C_D U_{10}^2, \quad (A1a)$$

$$Q_H = \rho c_p C_H U_{10} \Delta \theta_{10}, \quad (A1b)$$

$$Q_E = \rho L_V C_E U_{10} \Delta q_{10}, \quad (A1c)$$

where $\rho$ is air density; $C_D$, $C_H$, and $C_E$ are the exchange coefficients of the momentum, sensible heat, and latent heat, respectively; $c_p$ is the specific heat of humid air at a constant pressure; $L_V$ is the latent heat of vaporization; $\Delta \theta_{10} = \theta_0 - \theta_{10}$, and $\Delta q_{10} = q_0 - q_{10}$; $U_{10}$, $\theta_{10}$, and $q_{10}$ are the values of wind speed, temperature, and specific humidity at the 10 [m] reference height, respectively; $\theta_0$ and $q_0$ are the temperature and the specific humidity at the sea level. Then

$$\bar{\tau} = \frac{1}{S}\int_S \tau dS' \equiv \frac{1}{S}\left(\int_{S_w} \tau_w dS' + \int_{S_f} \tau_f dS'\right), \quad (A2a)$$

$$\bar{Q}_N = \frac{1}{S}\int_S Q_N dS' \equiv \frac{1}{S}\left(\int_{S_w} Q_{Nw} dS' + \int_{S_f} Q_{Nf} dS'\right), \quad (A2b)$$

where $N = H, E$; the overbars indicate the values averaged over a sea surface; $S_w$, $S_f$ and $S = S_w + S_f$ are the areas of the foam-free, foam-covered, and total sea surfaces, respectively. Then the transfer coefficients $\bar{C}_{D,N}$ in Eqs. (A1) can be split into the sums of their partial values over the foam-free and foam-covered sea surfaces weighted with the foam coverage coefficient, $\alpha_f$:

$$\bar{C}_{D,N} = (1 - \alpha_f)\bar{C}_{D,Nw} + \alpha_f \bar{C}_{D,Nf}. \tag{A3}$$

Here

$$\bar{\tau}_w = \rho \bar{C}_{Dw} U_{10}^2, \qquad \bar{\tau}_f = \rho \bar{C}_{Df} U_{10}^2, \qquad \bar{\tau} = \rho \bar{C}_D U_{10}^2, \tag{A4a}$$

$$\bar{Q}_{Hw} = \rho c_p \bar{C}_{Hw} U_{10} \Delta\theta_{10}, \quad \bar{Q}_{Hf} = \rho c_p \bar{C}_{Hf} U_{10} \Delta\theta_{10}, \quad \bar{Q}_H = \rho c_p \bar{C}_H U_{10} \Delta\theta_{10}, \tag{A4b}$$

$$\bar{Q}_{Ew} = \rho L_V \bar{C}_{Ew} U_{10} \Delta q_{10}, \quad \bar{Q}_{Ef} = \rho L_V \bar{C}_{Ef} U_{10} \Delta q_{10}, \quad \bar{Q}_E = \rho L_V \bar{C}_E U_{10} \Delta q_{10} \tag{A4c}$$

are the corresponding averaged values of momentum, heat, and moisture fluxes across the foam-free, foam-covered, and total sea surfaces, respectively; $\alpha_f = S_f/S$ is the foam coverage approximated as

$$\alpha_f = \gamma \tanh\left[\alpha \exp\left(\beta \frac{U_{10}}{U_{10}^{(S)}}\right)\right] \tag{A5}$$

with $\alpha = 0.00255$, $\beta \approx 8$, $\gamma = 0.98$, and $U_{10}^{(S)} \approx 48 \, [ms^{-1}]$. The above formula proposed by Holthuijsen et al. 2012 approximates observation data for $\alpha_f$ up to $U_{10} < 50 \, [ms^{-1}]$ and extended for $U_{10} > 50 \, [ms^{-1}]$ in GS2018.

According to the Monin-Obukhov similarity theory, the bulk transfer coefficients are related to atmospheric stability and reference height of measurements. To eliminate the influence of these factors, the bulk transfer coefficients are reduced to neutral stability atmospheric conditions with a standard 10 $[m]$ reference height (e.g. Andreas et al., 2012; Zou et al. 2017; Sergeev et al. 2017, and references therein). As it is noted in (Hsu 2003; Hsu et al. 2017), the logarithmic vertical profile of the mean wind speed can be adopted in real hurricane conditions under near-neutral stability atmospheric conditions. This results separately to the foam-free, foam-covered and total sea surfaces

$$\bar{C}_{Dw} = \frac{\kappa}{\ln[Z_{10}/Z_{Dw}]}, \quad \bar{C}_{Df} = \frac{\kappa}{\ln[Z_{10}/Z_{Df}]}, \quad \bar{C}_D = \frac{\kappa}{\ln[Z_{10}/Z_D]}, \tag{A6a}$$

$$\frac{\bar{C}_{Nw}}{\sqrt{\bar{C}_{Dw}}} = \frac{\kappa}{\ln[Z_{10}/Z_{Nw}]}, \quad \frac{\bar{C}_{Nf}}{\sqrt{\bar{C}_{Df}}} = \frac{\kappa}{\ln[Z_{10}/Z_{Nf}]}, \quad \frac{\bar{C}_N}{\sqrt{\bar{C}_D}} = \frac{\kappa}{\ln[Z_{10}/Z_N]}. \tag{A6b}$$

Obviously, the equations (A6) at known values of the transfer coefficient in their lefthand sides can be regarded as definitions of the roughness length parameters $\{Z_D, Z_{Dw}, Z_{Df}\}$ and $\{Z_N, Z_{Nw}, Z_{Nf}\}$ across the corresponding portions of the sea surfaces. In the present study, it is assumed that the transfer coefficients $\bar{C}_D$ and $\bar{C}_K$ across the total area of the alternating foam-free and foam-covered portions are estimated in the open-sea conditions. Whereas the transfer coefficients across the foam-free portions of the sea surfaces can be directly estimated in laboratory conditions as $\bar{C}_{D,N} = \bar{C}_{D,Nw}$, while the foam input in $\bar{C}_{D,N}$ is negligibly small. The transfer coefficients over the sea surface portion completely covered by foam $C_{D,Kf}$ cannot be measured directly, and the splitting relations (A3) can be used for their definition.

As commonly accepted now, the latent and sensible heat exchange coefficients have the same shape as the enthalpy coefficient $U_{10}$ (Zhang et al. 2008; Bao et al. 2011; Komori et al. 2018)

$$\bar{C}_K(U_{10}) = \bar{C}_H(U_{10}) = \bar{C}_E(U_{10}). \tag{A7}$$

Then the latent and sensible heat transfer coefficients can be represented by the enthalpy coefficient, and we can set $N = K$.

For further clarity, the bars over the corresponding variables are omitted.

**Appendix B**

Let us remind the recent ad-hoc modeling of the drag transfer across the air-sea interface under hurricane conditions proposed and developed by GS2016, and GS2018. According to Monin-Obukhov similarity theory (Foken 2006), the logarithmic vertical profile of the mean wind speed can be adopted for an atmospheric boundary layer under neutral stability conditions in typical hurricanes (Hsu 2003; Hsu et al. 2017)

$$U(Z) = \frac{U_*}{\varkappa} \ln\left(\frac{Z}{Z_D}\right), \tag{B1}$$

where $\varkappa = 0.4$ is von Karman's constant; $U_*$ [$ms^{-1}$] is the friction velocity; $Z$ [$m$] is the current height over the sea surface; $Z_D$ [$m$] is the effective aerodynamic roughness length over the total sea surface that consists of alternating foam-free and foam-covered fractions. Then the surface momentum flux $\tau = \rho C_D U_{10}^2$ provides the drag coefficient, $C_D$, that varies with the neutral stability wind speed $U_{10}$ at a reference height $Z_{10} = 10$ [$m$] ($\rho$ is the air density):

$$C_D = \frac{U_*^2}{U_{10}^2} \equiv \left(\frac{\varkappa}{\ln(Z_{10}/Z_D)}\right)^2. \tag{B2}$$

Conventionally, the problem of drag prediction is solved by specifying the effective roughness length $Z_D$. Thus, for relatively weak winds, the roughness length is well approximated by the well-known formula of Charnock 1955, $Z_D = \sigma_{Ch} U_*^2/g$, where $g$ is the gravity acceleration and $\sigma_{Ch}$ is a phenomenological constant. At negligibly small foam coverage, a better correspondence of the drag coefficient to the available experimental data in a wind speed range $U_{10} \leq 33\ [ms^{-1}]$ provides the model (Large and Pond 1981) modified, in accordance with the laboratory observations, by Donelan et al. 2004, by constant value of $C_D$ at $U_{10} > 33$ [ms$^{-1}$]:

$$C_D(U_{10}) = 10^{-3} \begin{cases} 1.14, & \text{for } 4\ [ms^{-1}] \leq U_{10} \leq 10\ [ms^{-1}], \\ 0.49 + 0.065 U_{10}, & \text{for } 10\ [ms^{-1}] \leq U_{10} \leq 33\ [ms^{-1}], \\ 2.6 & \text{for } 33\ [ms^{-1}] \leq U_{10}. \end{cases} \quad (B3)$$

GS2018 studies the foam input in the drag transfer coefficient in a wide range of $U_{10}$ from low to hurricane values. For this aim, the approximation (B3) has been adopted in GS2018. However, as mentioned in GS2018, $C_D \approx C_{Dw}$ for foam-free conditions can be estimated from any other laboratory experiments. In the GS2018 model, the drag transfer coefficient $C_D$ across the total sea-surface area of alternating foam-free and foam-covered fractions is estimated using the measurement data of Jarosz 2007 widely accepted as fiducial ones for the open-sea conditions. The resulting dependences $C_D$ and $C_{Dw}$ vs $U_{10}$ shown in Fig. 1 strongly differ one from another which is related to the foam presence on the sea surface. Assuming that $C_D(U_{10})$ and $C_{Dw}(U_{10})$ are observed in open-sea and laboratory conditions, respectively, Eq. (2) determines $Z_D(U_{10})$, and $Z_{Dw}(U_{10})$ is found from the following relation

$$C_{Dw} = \frac{U_{*w}^2}{U_{10}^2} \equiv \left(\frac{\varkappa}{ln(Z_{10}/Z_{Dw})}\right)^2. \quad (B4)$$

Then $Z_D$ is presented as an ad-hoc sum of two aerodynamic roughness lengths for the foam-free and foam-covered surfaces $Z_{Dw}$ and $Z_{Df}$ weighted by the fractional foam coverage, $\alpha_f$ (GS2016):

$$Z_D = (1 - \alpha_f) Z_{Dw} + \alpha_f Z_{Df}, \quad (B5)$$

where $\alpha_f(U_{10})$ is the foam coverage (Appendix A). Whereas $Z_D(U_{10})$ and $Z_{Dw}(U_{10})$ are found from relations (B2) and (B4) using the measured values $C_D$ and $C_{Dw}$, $Z_{Df}$ is determined from relation (B5). Dependencies $Z_{Dw}$ and $Z_{Df}$ vs $U_{10}$ adopted from GS2018 are shown in Fig. 2.